\documentclass[aps,prl,twocolumn,superscriptaddress]{revtex4-1}

\newcommand {\etal}{{\it et} {\it al.}}
\newcommand {\rxx}{\rho _{xx}}
\newcommand {\ryx}{\rho _{yx}}
\newcommand {\sxx}{\sigma _{xx}}
\newcommand {\sxy}{\sigma _{xy}}
\newcommand {\ef}{E_{\rm F}}
\newcommand {\tn}{T_{\rm{N}}}
\newcommand {\mW}{\rm{m\Omega\ cm}}

\newcommand {\rio}{R_{2}{\rm Ir}_{2}{\rm O}_{7}}
\newcommand {\gcio}{({\rm Gd}_{1-x}{\rm Cd}_{x})_{2}{\rm Ir}_{2}{\rm O}_{7}}
\newcommand {\ecio}{({\rm Eu}_{1-x}{\rm Ca}_{x})_{2}{\rm Ir}_{2}{\rm O}_{7}}
\newcommand {\tcio}{({\rm Tb}_{1-x}{\rm Ca}_{x})_{2}{\rm Ir}_{2}{\rm O}_{7}}

\usepackage[dvipdfmx]{graphicx}
\usepackage{bm}
\usepackage{pifont}
\usepackage{amsmath,amssymb}
\usepackage[dvipdfmx]{color}

\begin{document}

\title{Evolution of possible Weyl semimetal states across the hole-doping induced Mott transition in pyrochlore iridates}

\author{K. Ueda}
\affiliation{Department of Applied Physics, University of Tokyo, Tokyo 113-8656, Japan}

\author{H. Fukuda}
\affiliation{Department of Applied Physics, University of Tokyo, Tokyo 113-8656, Japan}

\author{R. Kaneko}
\affiliation{Department of Applied Physics, University of Tokyo, Tokyo 113-8656, Japan}
\affiliation{RIKEN Center for Emergent Matter Science (CEMS), Wako 351-0198, Japan}

\author{J. Fujioka}
\affiliation{Faculty of Material Sciences, University of Tsukuba, Tsukuba 305-8577, Japan}

\author{Y. Tokura}
\affiliation{Department of Applied Physics, University of Tokyo, Tokyo 113-8656, Japan}
\affiliation{RIKEN Center for Emergent Matter Science (CEMS), Wako 351-0198, Japan}
\affiliation{Tokyo Colledge, University of Tokyo, Tokyo 113-8656, Japan}

\date{\today}

\begin{abstract}
We study possible Weyl semimetals of strongly-correlated electrons by investigating magnetotransport properties in pyrochlore $R_2$Ir$_2$O$_7$ ($R$=rare-earth ions), choosing three types of $R$ ions to design the exchange coupling scheme between $R$ $4f$ and Ir $5d$ moments; non-magnetic Eu ($4f^6$), isotropic Gd ($4f^7$), and anisotropic Tb ($4f^8$).
In the doping-induced semimetallic state, distinctive features of magnetoresistance and Hall effect are observed in $R$=Gd and Tb compounds due to the effects of the exchange-enhanced isotropic and anisotropic Zeeman fields, respectively, exemplifying the double Weyl semimetal and the 2-in 2-out line-node semimetal as predicted by theories.
In particular, a Hall angle of $R$=Gd compound is strongly enhanced to 1.5 \% near above the critical doping for the Mott transition.
Furthermore, an unconventional Hall contribution is discerned for a lower doping regime of $R$=Gd compound, which can be ascribed to the emergence of Weyl points with the field-distorted all-in all-out order state.
These findings indicate that the hole-doping induced Mott transition as well as the characteristic $f$-$d$ exchange interaction stabilizes versatile topological semimetal states in a wide range of material parameter space.
\end{abstract}

\maketitle

\noindent

\section{Introduction}
The correlation between magnetism and topological electronic states has been one of the most important themes of the modern condensed matter physics \cite{2018RMPArmitage,2019NatureTokura}.
A magnetic Weyl semimetal (WSM) possesses crossings of non-degenerate bands with linear dispersion under the breaking of time-reversal symmetry with magnetic order \cite{2011PRBWan}.
Notably, the crossing (Weyl point) can be viewed as a magnetic monopole of Berry curvature in the momentum space, proven to yield salient magnetotransport properties \cite{2011PRBZyuzin}.
Prominent example is the intrinsic anomalous Hall effect (AHE) \cite{2010RMPNagaosa}.
Recent enormous efforts on this subject reveal that the position of Weyl points relative to the Fermi energy $\ef $ is intimately related to the size of AHE, revealing astoundingly large AHE in a variety of magnets \cite{2015NatureNakatsuji,2016NPhysSuzuki,2018NPhysLiu,2018NCommGhimire}.
Despite such remarkable findings, the role of electron-correlation as a source of magnetism in topological electronic states remains elusive.
In particular, the emergence of the magnetic topological states over the electron-correlation induced metal-insulator transition (MIT) \cite{1998RMPImada} has seldom been explored.

Pyrochlore oxide $A_2B_2$O$_7$ is known to host exotic quantum magnetic ground states \cite{2010RMPGardner}, and recently proposed to host topological electronic states for iridium oxides $R_2$Ir$_2$O$_7$ ($R$ being rare-earth or Y ions) \cite{2011PRBWan}.
The merit of pyrochlore oxides is that the effective electron correlation $U/W$ and the band filling $n$ (=1-$\delta $, $\delta $ being hole-doping) can be tuned by varying $A$-site ions in a similar manner to perovskites \cite{1998RMPImada}.
For instance, $R_2$Ir$_2$O$_7$ with a smaller ionic-size $R$ shows an effectively larger bending of Ir-O-Ir bond angle and hence leads to a smaller one-electron band-width $W$; thus the $R$-ion size is a good indicator of the inverse of the effective electron correlation $U/W$, as shown in Fig.1(a).
Among a variety of $R_2$Ir$_2$O$_7$, $R$=Pr compound, which is characterized by the largest $R$ ionic radius, is found to possess a quadratic band touching (QBT) of spin degenerate valence and conduction bands near $\ef $ in the paramagnetic metal (PM) phase (see Fig. 1(b)) \cite{2015NCommKondo}.
Since the QBT is protected by the cubic crystalline symmetry and the time-reversal symmetry, it is expected to be robust against the perturbation as far as the symmetry is kept intact.
Such an electronic state is termed Luttinger SM or QBT SM that is theoretically anticipated to be converted to versatile topological states by symmetry-breaking procedures \cite{2013PRLMoon,2016PRBCano}.
Figure 1(b) displays a schematic picture of the electronic band modulation across Mott transitions.
The strong Hubbard repulsion $U$ brings about the antiferromagnetic-like all-in all-out (AIAO) type magnetic order (depicted in the inset in Fig. 1(a)), and lifts the band degeneracy of QBT, leading to the emergence of WPs along the high-symmetry axes \cite{2018RMPArmitage,2012PRBKrempa}, termed AIAO WSM hereafter.
Although WSM is generally robust against the perturbation, WPs in this system emigrate towards the Brillouin zone boundary with increasing $U/W$ or the magnetic order parameter $m$. Consequently, with a tiny increase of $U$ or $m$, WSM undergoes the pair annihilation of WPs at the zone boundary and turns into a gapped insulating state (Fig. 1(b)), hampering the observation in the momentum space.
In spite of such difficulty, intensive studies on $R$=Nd and Pr compounds have unveiled a number of emergent transport phenomena such as unusual Hall effect \cite{2010NatureMachida,2018NCommUeda}, magnetic field-induced MITs \cite{2015PRLUeda,2016NPhysTian,2017NCommUeda}, and anomalous metallic states on the magnetic domain walls in the AIAO state\cite{2014PRBUeda,2015ScienceMa}, which are all understood in the context of underling or incipient WSM sttaes induced by the time-reversal symmetry breaking (Fig. 7 in Appendix).

On the other hand, $R_2$Ir$_2$O$_7$ with smaller $R$ ionic radius are not QBT-SM but fully-gapped insulators because of larger $U/W$ (Fig. 1(a)), and thus have been considered to be no longer related to topological semimetallic state \cite{2016PRBUeda}.
However, recent studies demonstrate that the hole-doping by the chemical substitution of trivalent $R$ ions with divalent $A$ ions turns them into PM state\cite{2002JPSJFukazawa,2014PRBZhu,2019PRBKaneko,2019PRBPorter} in which, remarkably, QBT-SM can ubiquitously subsist even if the chemical substitution introduces some disorders \cite{2019PRBKaneko}
In this sense, the hole-doped $\rio $ offers a fertile playground to study the correlation between magnetism and topological electronic states, since carrier doping can tune the position of the QBT node with respect to $\ef $, and moreover, a variety of local $R$-$4f$ magnetic spins, which strongly affect itinerant Ir-$5d$ electrons through the $f$-$d$ coupling, are available in this pyrochlore system.

\section{Results}
\subsection{Hole-doping induced insulator-metal transition}
In this work, we choose three $R$ ions to study the role of $R$ magnetic moment in the magnetic-field induced modification of QBT state; nonmagnetic $R$=Eu ($4f^6$), isotropic $R$=Gd ($4f^7$), and  Ising-type $R$=Tb ($4f^8$).
We synthesized high-quality polycrystals by utilizing high-pressure apparatus which promotes the pyrochlore-lattice formation while keeping the right stoichiometry of compounds.
Employing the growth condition described elsewhere \cite{2012PRLUeda}, we obtained the hard and dense samples suitable for the systematic transport measurements.
As for the chemical substitution of trivalent $R$ ions with divalent $A$ ions, we found that $A$=Ca (Cd) is suitable for $R$=Eu and Tb (Gd) in the light of the ion-size matching.
It allows us to examine the hole doping effect while suppressing other contributions such as the $A$-site doping-induced change of bandwidth or disorder. Zhu {\etal} demonstrate that the Ca-doping to Y$_2$Ir$_2$O$_7$ enhances both metallicity and ferromagnetism \cite{2014PRBZhu}. They found that the temperature dependence of resistivity shows a minimum at around 100 K for the high-doped metallic sample, which is somewhat different from our result as shown below. We speculate that it is due to the mismatch of $A$-site ionic radius. Since the Ca ionic radius is much larger than that of Y ions, the Ca doping modifies the effective bandwidth via the change of Ir-O-Ir bond angles, and thereby brings about the resistivity minimum at around 100 K, which is also observed in Eu$_2$Ir$_2$O$_7$ under the hydrostatic pressure above 7.88 GPa \cite{2012PRBTafti}.

As the $A^{2+}$ concentration $x$ increases in $\gcio $ and $\tcio $, which corresponds to the nominal hole-doping $\delta $, both exhibit the systematic reduction in resistivity $\rxx $ as well as in $\tn $ that shows up as a kink of $\rxx $ in accord with the anomaly in $M$. They turn into PM at sufficiently large $x$, although the resistivity slightly increases below 20 K as shown in Fig. 2.
The results including $\tn $ are summarized in the phase diagram, Fig. 1(a).
Incidentally, the upturn of $\rxx $ at low temperatures in $\tcio $ and $\gcio $, which is absent in $\ecio $, is due perhaps to the strong magnetic coupling between itinerant Ir electrons and localized $R$ moments which produces incoherent carrier-electron scattering \cite{2012PRLUdagawa,2016PRLWang}.
Since $\rxx $ value of (Eu$_{0.90}$Ca$_{0.10}$)$_2$Ir$_2$O$_7$ is close to those of (Gd$_{0.88}$Cd$_{0.12}$)$_2$Ir$_2$O$_7$ and (Tb$_{0.85}$Ca$_{0.15}$)$_2$Ir$_2$O$_7$, the similar electronic states that host QBT at $\Gamma $ point are anticipated to be realized in hole-doped analogs of both $R$=Gd and Tb compounds, which is corroborated by magnetotranport measurements described in the following.

\begin{figure}
\begin{center}
\includegraphics[width=3.4in,keepaspectratio=true]{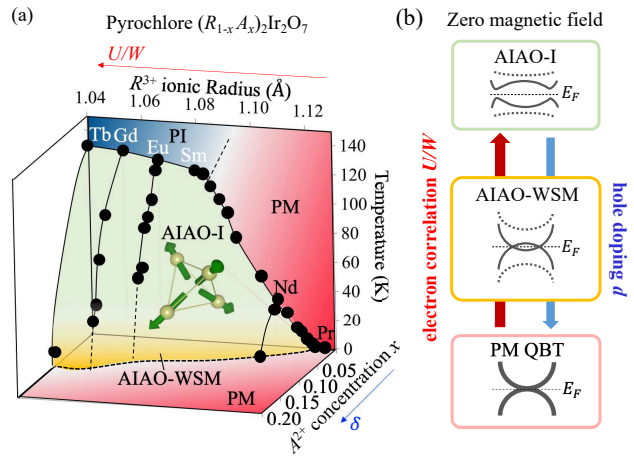}
\caption{(color online).
(a) Phase diagram of $R_2$Ir$_2$O$_7$ as functions of rare-earth ionic radius, divalent-ion concentration, and temperature. PI stands for paramagnetic insulator, PM stands for paramagnetic metal, AIAO stands for all-in all-out state, and WSM stands for Weyl semimetal.
(b) Schematic picture of modulation of electronic band structures as a function of electron correlation $U$ and hole doping $\delta $. QBT stands for quadratic band touching.
}
\end{center}
\end{figure}

\begin{figure}
\begin{center}
\includegraphics[width=2.7in,keepaspectratio=true]{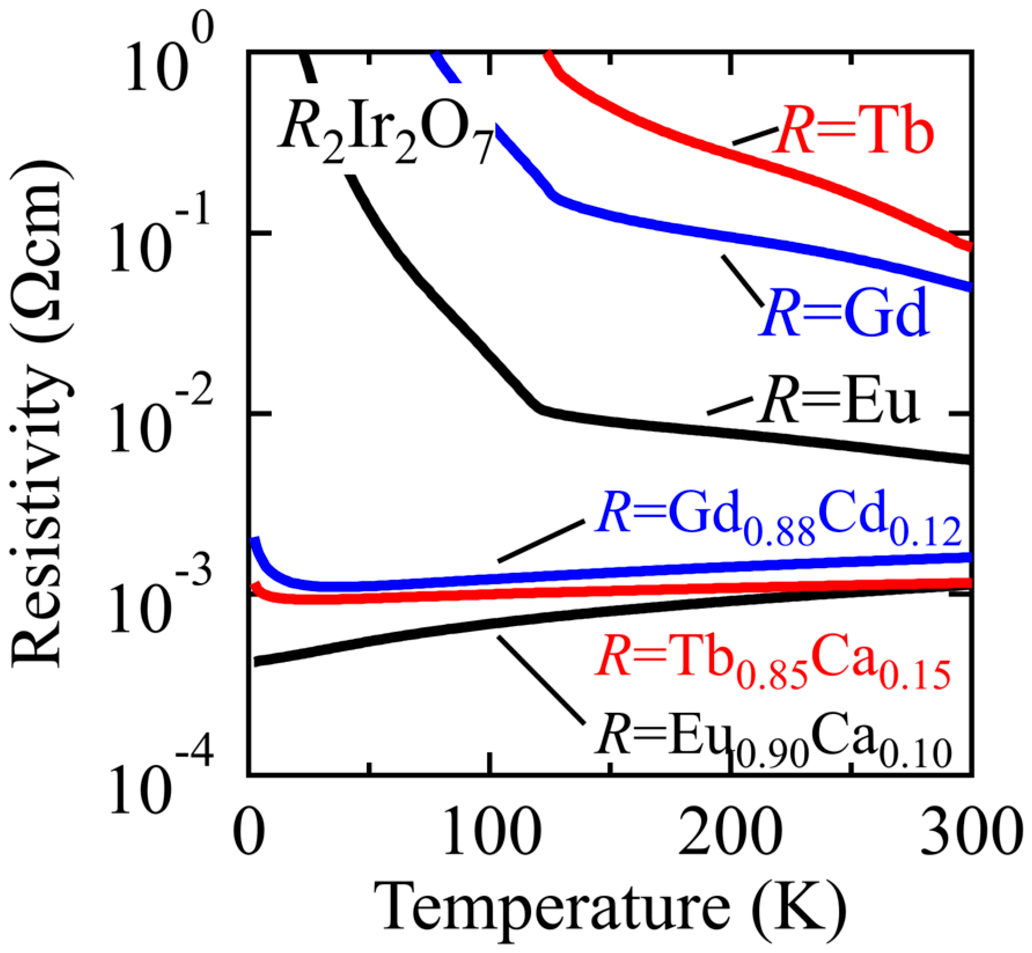}
\caption{(color online).
Temperature dependence of resistivity for $\ecio $ ($x$=0, 0.10), $\gcio $ ($x$=0, 0.12), and $\tcio $ ($x$=0, 0.15).
}
\end{center}
\end{figure}

\begin{figure*}
\begin{center}
\includegraphics[width=4.5in,keepaspectratio=true]{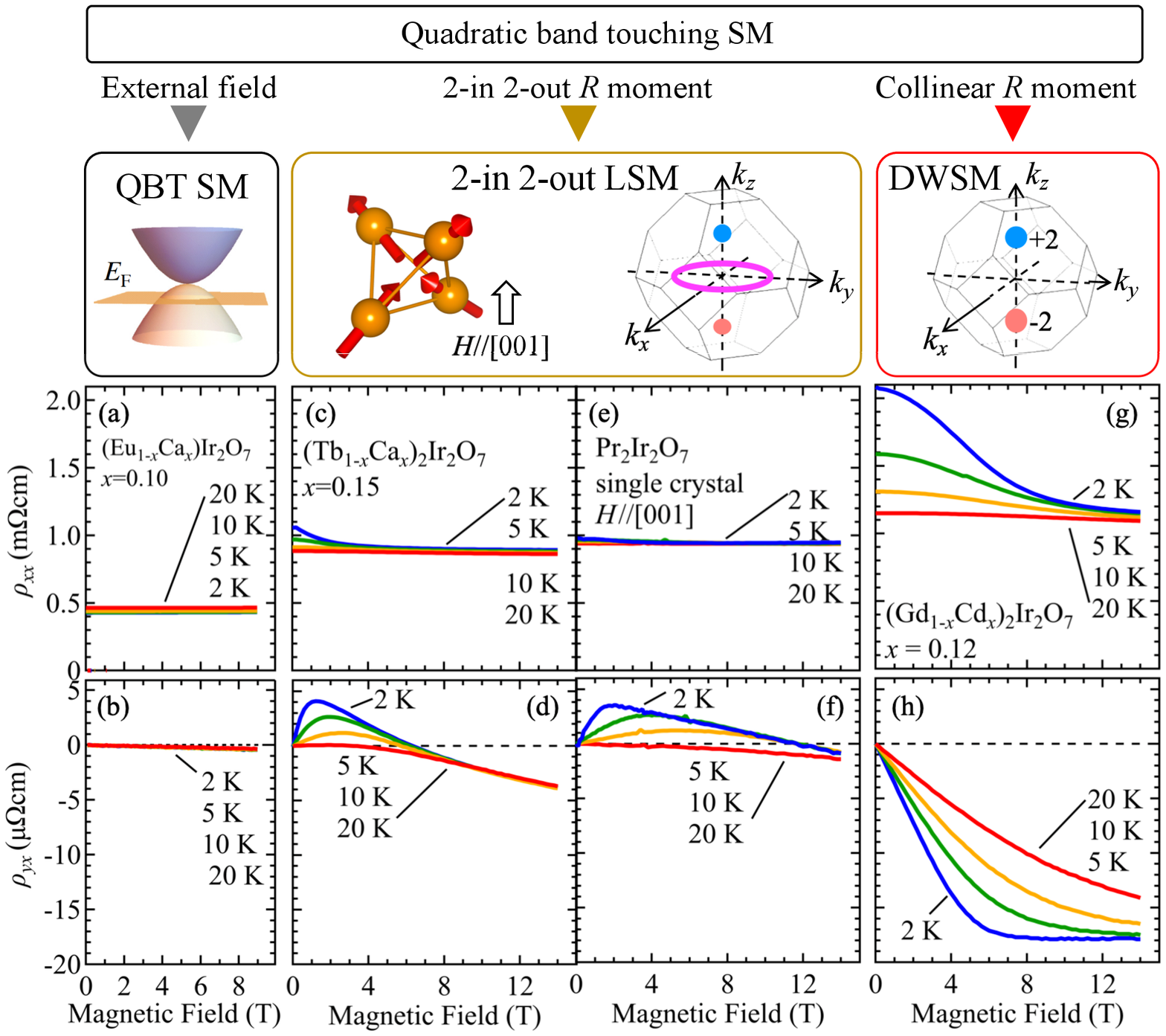}
\caption{(color online). 
Magnetic field dependence of longitudinal resistivity ((a),(c),(e),(g)) and Hall resistivity ((b),(d),(f),(h)). From left to right, the data are for (a,b) $\ecio $ ($x$=0.10), (c,d) $\tcio $ ($x$=0.15), (e,f) Pr$_2$Ir$_2$O$_7$. (c,d) $\gcio $ ($x$=0.12), respectively. Schematic magnetic configurations of $R$ spins and expected electronic structures are displayed on top of respective windows. 
}
\end{center}
\end{figure*}

\subsection{Magnetotransport of metallic $R$ compounds}
Having thus realized the QBT state in various magnetic ($R_{1-x}A_{x}$)$_2$Ir$_2$O$_7$, we study the magnetic field dependence of $\rho _{xx}$ and Hall resistivity $\rho _{yx}$ in Fig. 3.
(Eu$_{0.90}$Ca$_{0.10}$)$_2$Ir$_2$O$_7$, whose $R$-site ion is non-magnetic, shows little magnetic field dependence of $\rxx $ (Fig. 3(a)) and $\ryx $ apart from the normal Hall effect (Fig. 3(b)).
On the other hand, negative magnetoresistance is clearly observed for $\tcio $ with $x$=0.15 in Fig. 3(c), indicating the coupling between the conduction electron and $R$-ion moment. More importantly, $\ryx $ shows a complicated magnetic-field profile especially at 2 K (Fig. 3(d)); as the field increases, $\ryx $ abruptly increases with a maximum around 1 T, and subsequently plunges to the negative value, changing the sign at 7 T.
To identify the origin of non-monotonic Hall responses for $\tcio $, we examine those of well-oriented single crystal Pr$_2$Ir$_2$O$_7$ which is well characterized by QBT SM state at zero field and zero doping ($\delta =0$) \cite{2015NCommKondo}, and an Ising anisotropy of $R$ moment similar to the Tb one.
Figure 3(e) and 3(f) display $\rxx $ and $\ryx $ of Pr$_2$Ir$_2$O$_7$ (single crystal) respectively, as a function of the magnetic field along [001] axis which favors 2-in 2-out state as depicted on top of Figs.3 (c), (e). Both $\rxx $ and $\ryx $ are qualitatively similar to those of $\tcio $ ($x$=0.15). A former study combining transport measurements on (Nd,Pr)$_2$Ir$_2$O$_7$ and theoretical calculation suggests that the magnetic field can modulate the electronic state into the line-node SM (LSM) depicted in the middle panel of Fig.2 via the magnetic stabilization at the 2-in 2-out state \cite{2015PRLUeda}. Thus, the close similarity of $\rxx $ and $\ryx $ in the two compounds with similar Ising anisotropy of $R$ moments implies that the LSM can be also realized in $\tcio $ polycrystals at high fields where the 2-in 2-out configuration is dominant.

On the other hand, a very different trend of $\rxx $ and $\ryx $ is observed in $\gcio $ ($x$=0.12).
Especially, at the lowest temperature 2 K, $\rxx $ markedly decreases by half at 14 T (Fig. 3(g)) while the absolute value of negative $\ryx $ increases rapidly. This anomalous Hall-like behavior in the originally paramagnetic but field-magnetized state can show the large $\ryx $ value, 20 times larger than that of $R$=Eu compound (Fig. 3(h)), and the observed value of the Hall angle is as large as 1.5 $\% $. 
For instance, the Hall angle of ferromagnetic oxides such as SrRuO$_3$ \cite{2003ScienceFang} and (La,Sr)CoO$_3$ \cite{2007PRLMiyasato} is approximately 0.9 $\% $, and that of isostrural ferromagnet Nd$_2$Mo$_2$O$_7$ with the 2-in 2-out scalar spin chirality is about 1.3 $\% $ \cite{2007PRLIguchi}.
Such a large Hall effect can point to a substantial Berry-curvature contribution near $\ef $ in the momentum space.
Oh $et$ $al.$ theoretically show that the uniform magnetic field along [001] crystalline direction lifts the band degeneracy into two Weyl points hosting the monopole charge $\pm $2 on [001] momentum axis, as depicted on top of Fig.3 (g) (and in Fig. 7, Appendix) \cite{2018PRBOh}. It is termed double-Weyl semimetal (DWSM).
Apparently, this situation cannot be caused by a simple external-field-induced Zeeman splitting; if so, the $R$=Eu compound would show the similar effect, but not in reality (Fig. 3(b)).
The sufficiently large exchange splitting of QBT is only driven by the $f$-$d$ exchange coupling from six nearest-neighbor Gd spins which are easily aligned by external fields (Fig. 4(a)).

\begin{figure*}[htb]
\includegraphics[width=4.5in,keepaspectratio=true]{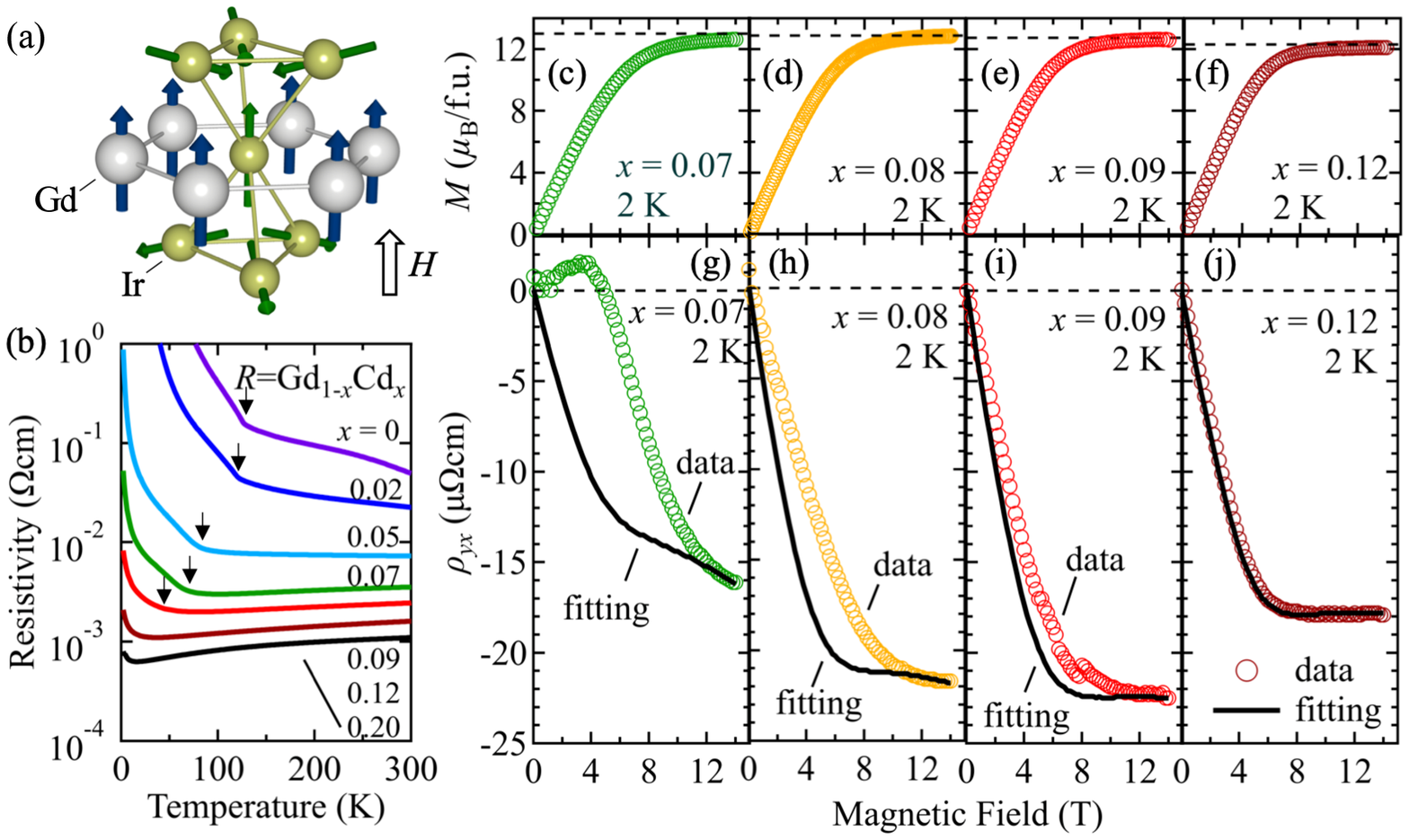}
\caption{(color online). 
(a) Schematic magnetic structures of Ir moments (green) and Gd moments (blue).
(b) Temperature dependence of resistivity for $\gcio $ with various $x$. Arrows indicate the magnetic transition temperature.
Magnetic field dependence $H$ of magnetization $M$ ((c),(d),(e),(f)) and Hall resistivity ((g),(h),(i),(j)) for various $x$ in $\gcio $. From left to right, the data are for (c,g) $x$=0.07, (d,h) $x$=0.08, (e,i) $x$=0.09, (f,j) $x$=0.12, respectively. Thick black lines are fitting curves with the $H$-linear normal Hall plus the $M$-linear anomalous Hall terms (see text).
}
\end{figure*}

\subsection{Magnetotransport across the filling-control-induced metal-insulator transition in $\gcio $}
In order to gain deeper insight into the possible topological-state change in $\gcio $, we implement a precise investigation on magnetotransport properties across doping-induced Mott transition.
Figure 4(b) shows the temperature dependence of $\rxx $ for several $\gcio $ compounds with varying $x$. Accompanied by the decline of $\rxx $, $\tn $, which is indicated by arrows in Fig. 4(b), is gradually suppressed with increasing $x$ and no longer discernible for $x>0.10$.
The temperature dependence of $M$ for several $x$ is displayed in Fig. 8, Appendix. At $x$=0.07, $M$ in the field-cooling process clearly deviates from that in the zero-field-cooling process below $\tn $ at which $\rxx $ shows a kink. The deviation disappears down to 2 K for $x$=0.12, although the upturn of $\rxx $ subsist. The upturn of $\rxx $ is also observed in Ca-doped Nd$_2$Ir$_2$O$_7$ below 20 K \cite{2019PRBPorter}, and attributed to the ordering or freezing of the Nd magnetic sublattice which is observed by muon spin resonance \cite{2013PRBGuo}.
Figures 4(c-j) show the magnetic field dependence of $M$ and $\ryx $ at 2 K for various compositions near the Mott transition, ranging from the $x$=0.07 compound, which undergoes long-range AIAO order below $\tn $, to the paramagnetic semimetal $x$=0.12 compound.
As shown in Figs. 4(c)-(f), $M$ increases monotonically with increasing field and finally saturates at high fields for all compositions. The saturated $M$ values are consistent with those expected from fully-aligned Gd spins as depicted by dashed lines. The critical magnetic field, which is required to reach the saturated $M$, gradually becomes smaller as $x$ increases; it is roughly 11 T for $x$=0.07 while 8 T for $x$=0.12. It is presumably because the AIAO-type antiferromagnetic correlation of Ir moments in the small $x$ region competes with the collinear alignment of Gd moments via the $f$-$d$ coupling.
Figures 4 (g-j) display $\ryx $ at 2 K as a function of the magnetic field. Interestingly, $\ryx $ for $x$=0.07 shows a nonmonotonic field dependence that is explicitly distinguished from the usual AHE (Fig. 4(g)). As the field increases, $\ryx $ increases with a broad maximum around 3 T, and then markedly decreases with the sign reversal. Finally $\ryx $ decreases modestly above 11 T at which $M$ is saturated. This characteristic field dependency at 2 K fades out into the monotonic change as the temperature is elevated (see Fig. 9 in Appendix).
On the other hand, $\ryx $ for $x$=0.12 monotonically decreases and saturates at around 8 T, seemingly in proportion to $M$ (Fig. 4(j)).

\begin{figure}
\begin{center}
\includegraphics[width=2.8in,keepaspectratio=true]{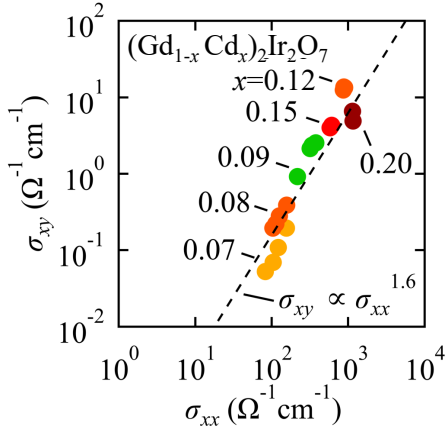}
\caption{(color online).
Log-log plot of Hall conductivity versus longitudinal conductivity for $\gcio $ with various $x$. The dashed line indicates the relation $\sxy \propto \sxx ^{1.6}$.}
\end{center}
\end{figure}

To understand this behavior, we employ the fitting function which is conventionally used for ferromagnets, expressed as $\ryx ^{\rm fit}=R_{0}B+\mu _{0}R_{s}M$, where $R_{0}$ is the normal Hall coefficient and $R_{s}$ is the anomalous coefficient.
The Hall conductivity $\sxy $ values for 0.07 $\leq $ $x$ $\leq $ 0.20 taken at 14 T, i.e. presumably in the DWSM state, fairly well satisfy the relation that $\sxy \propto \sxx ^{1.6}$ (Fig. 5), which is empirically known for the intrinsic anomalous Hall effect in the large-scattering regions, e.g. $\sigma _{xy}<$ 10$^4$ S/cm \cite{2008PRBOnoda}.
Conversely, this indicates the common origin, i.e. the anomalous Hall effect characteristic of DWSM, for the high-field $\sxy $ irrespective of $x$ or $\sxx $ value.
Thus we use $R_{s}=S_{\rm H}\rho _{xx}^{0.4}$ with the scaling coefficient $S_{\rm H}$.
The solid lines in Figs. 4 are fitting curves $\ryx ^{\rm fit}$ which reproduce $\ryx ^{\rm exp}$ in the field-induced Gd-spin collinear region, i.e. at high magnetic fields.
While $\ryx ^{\rm exp}$ of $x$=0.12 is well fitted with the above relation over the entire magnetic field range, those of other compounds in Figs. 4, which take AIAO states at zero field, show clear deviations from the fitting curves at low fields.
The deviation becomes larger as temperature decreases (Fig. 10, Appendix).
The observed $\ryx $ should reflect the variation of the electronic states through the change of the magnetic configurations.
At low fields, Ir spins form the AIAO type magnetic ordering configuration. As the field increases, Gd spins gradually point to the field direction and force Ir spins to align in a parallel manner through the $f$-$d$ exchange interaction, resulting in the breaking of the AIAO pattern accompanied with the remarkable change of $\rxx $ and $\ryx $ at high fields.
In this regard, the origin of the deviation can be ascribed to the emergence of WPs with "distorted" AIAO magnetic state.
Note here that the AIAO maintains the cubic crystalline symmetry, and hence the net Berry curvature integrated over the Brillouin zone is canceled out in the ideal AIAO WSM. However, when the Weyl vectors connecting a pair of WPs are deformed by the Zeeman field, the Hall response can show up to a considerable value, as demonstrated in Nd$_2$Ir$_2$O$_7$ under pressure \cite{2018NCommUeda} or epitaxially-strained film \cite{2020SciAdvKim}.
Hereafter we define the deviating part $\Delta \ryx =\ryx ^{\rm exp}-\ryx ^{\rm fit}$  as a contribution of Berry curvature in AIAO WSM.

\begin{figure}
\begin{center}
\includegraphics[width=2.3in,keepaspectratio=true]{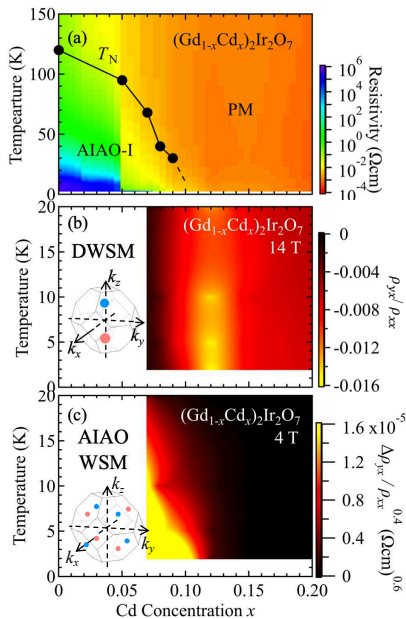}
\caption{(color online). 
(a) Phase diagram of $\gcio $ with contour mapping of resistivity in the plane of temperature and Cd concentration $x$. Black marks denote the magnetic transition temperature $\tn $.
Contour plots of (b) Hall angle $\ryx /\rxx $ at 14 T and (c) $\Delta \ryx $/$\rxx ^{0.4}$ at 4T, which represent the Hall responses from DWSM and AIAO WSM, respectively. 
}
\end{center}
\end{figure}

\section{Discussion}
To clarify the nature of the observed Hall effect in $\gcio $, we summarize the comprehensive results of $\rxx $ and $\ryx $ in contour mappings along with the phase diagram in Fig. 6.
We firstly show the phase diagram with the contour plot of $\rxx $ in the plane of temperature and Cd concentration $x$ in Fig. 6(a).
As $x$ increases, $\tn $ systematically decreases and disappears at around $x$=0.10. Concomitantly, $\rxx $ significantly decreases by several orders of magnitude down to $\sim $1 $\mW $  in the PM phase as discussed in Fig. 4(b).
Figures 6(b) and 6(c) display the contour plot of the Hall angle $\ryx /\rxx $ at 14 T and the $\Delta \ryx $ scaled with $\rxx ^{0.4}$ at 4 T, which are anticipated to represent the Hall responses from the DWSM and the AIAO WSM state, respectively.
Both signals are enhanced at low temperatures, suggesting that the magnetic coupling between Gd-$4f$ and Ir-$5d$ moments plays a vital role in the magnetotransport.
As $x$ increases, the Hall angle $\ryx /\rxx $ at 2 K and 14 T increases steeply, reaches a maximum of 1.5 $\% $ at $x$=0.12, and mildly decreases yet sustains 0.7 $\% $ at $x$=0.20 (Fig. 6(b)), which is far beyond the value of $\tcio $ which exhibits 2-in 2-out LSM.
Such a large Hall effect is attributable to efficient Berry-curvature generation in DWSM phase that host a pair of WPs with monopole charges of $\pm $2.
In addition, it can be also augmented by the strong magnetic interaction between conducting electrons and local moments that can manifest itself as the upturn of $\rxx $ (Fig. 2).
On the other hand, a finite value of $\Delta \ryx /\rxx ^{0.4}$ at 4 T (Fig. 6(c)) extends over the large area of AIAO phase ($x<$ 0.10).
It is in sharp contrast to the AIAO WSM phase at zero field for the undoped case of (Nd,Pr)$_2$Ir$_2$O$_7$, which is constricted in the narrow temperature window, i.e., within $\sim $2 K just below $\tn $ \cite{2018NCommUeda}.
According to theoretical calculation \cite{2018PRBOh}, the WPs in AIAO WSM move away from [111] or equivalent high-symmetry axes under the uniform Zeeman field which breaks threefold rotation symmetry. Consequently the pair annihilation is prevented in the broad parameter region, leading to the expansion of WSM phase as discussed for the case of (Nd,Pr)$_{2}$Ir$_{2}$O$_{7}$ \cite{2017NCommUeda}.
In short, as seen in Fig. 6, the compounds undergo the critical transformation from the AIAO WSM  to the QBT SM at zero field as well as to the DWSM under the fully aligned Gd moments, as the hole doping proceeds across the Mott transition point ($x \sim $0.10).
Here we note that the divalent ion substitution introduces some disorders to the system and could affect the topological electronic states, although the topological system is basically robust against them. In fact, Ref. \cite{2019PRBKaneko} demonstrates that the QBT state, which is realized in the presence of cubic symmetry, exhibit unique Seebeck effect even in the heavily doped compounds. Thus we speculate that the topological states including magnetic ones sustain in the doped systems as well. The robustness for the topological state against chemical disorder is an interesting question and remains to be solved in the future study.

\section{Summary}
We investigate the magnetotransport properties in the course of hole-doping induced Mott transitions for pyrochlore iridates $R$=Eu, Gd, and Tb.
We establish the versatile phase diagrams of the topological states, including the all-in all-out Weyl semimetal, the 2-in 2-out line-node semimetal, the quadratic-band-touching semimetal and the double-Weyl semimetal, as functions of $R$ ionic radius, temperature, magnetic field, and the band filling (hole doping).
Among them, $\gcio $ exhibit characteristic Hall effects which point to the different Berry-curvature generations in the two distinct topological semimetals, i.e. the field-distorted all-in all-out Weyl semimetal and the double Weyl semimetal.
The present work shows that the control of electron correlation by tuning not only the bandwidth but also the band filling can unravel the hidden topological semimetal state or dramatically expand its stable region in pyrocholore iridates where the field selection of the specific Weyl semimetals (WSMs) is also possible via the exchange coupling between the rare-earth $4f$ and Ir $5d$ moments.

\section{Acknowledgement}
We thank Hiroaki Ishizuka for enlightening discussions.
This work was supported by JSPS Grant-in-Aid for Scientific Research (No. 19K14647), and CREST (No. JPMJCR16F1 and JPMJCR1874), Japan Science and Technology Japan.

\section{Appendix}

\subsection{Possible topological semimetal states with different magnetic configurations}
The electronic state of pyrochlore iridates in the paramagnetic metal phase is predicted to host a quadratic band touching (QBT) at $\Gamma $ point which is actually observed by angle-resolved photoemission spectroscopy\cite{2015NCommKondo}.
Such an electronic state is anticipated to be converted to versatile topological states by symmetry breaking, as predicted in some theories \cite{2011PRBWan,2013PRLMoon,2017NCommUeda,2018PRBOh,2017PRBGoswami,2020arXivLadovrechis}.
Figure 7 displays a representative example for the electronic band modulation with magnetic ordering patterns.
In the case of 2-in 2-out type magnetic configuration, the line-node semimetal (LSM), which possesses a pair of Weyl points (WPs) along [001] direction and a line node in $k_z$ plane, can be stabilized.
On the other hand, three pairs of WPs show up when the 3-in 1-out type magnetic pattern is realized.
Furthermore, the exchange field of collinearly-aligned $R$ spins (or extremely large Zeeman field) can give rise to the double Weyl semimetal that host a pair of WPs with monopole charge of $\pm $2.

\begin{figure}
\begin{center}
\includegraphics[width=2.7in,keepaspectratio=true]{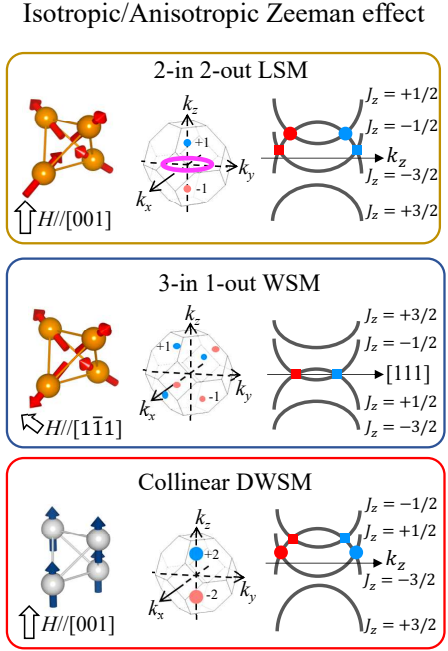}
\caption{(color online).
Schematic picture of electronic band structures with different magnetic ordering patterns stemming from the quadratic band touching semimetal at zero magnetic field in Fig. 1(b). From top to bottom, line-node semimetal (LSM) with 2-in 2-out magnetic state, Weyl semimetal (WSM) with 3-in 1-out magnetic state, and double Weyl semimetal (DWSM) with collinear magnetic state.
On the left side, respective magnetic ordering configuration is shown.
The middle column display the distribution of Weyl points and a line node in the Brillouin zone. The red and blue circles denote Weyl points with different chiralities and the purple line denotes a line node.
On the right side, we show the schematic band energy levels for respective electronic states. Red and blue squares (circles) denote Weyl points with monopole charges of $\pm $1 ($\pm $2).
}
\end{center}
\end{figure}

\section{Transport and magnetization properties in $\gcio $}
Figure 8 shows the temperature dependence of resistivity and magnetization for various $\gcio $. For $x$=0.07, the magnetization in field-cooling process (red lines) clearly deviates from that in zero-field-cooling process (blue line) at the magnetic transition temperature, at which the resistivity shows slight upturn. As $x$ increases, the transition temperature systematically decreases while the resistivity decreases. The anomaly is no longer observed for $x$=0.12.

\begin{figure}
\begin{center}
\includegraphics[width=3.5in,keepaspectratio=true]{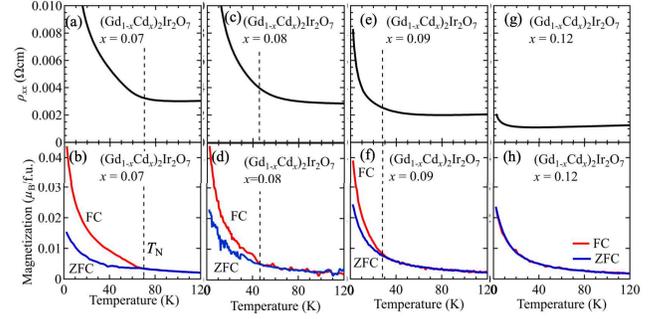}
\caption{(color online).
Temperature dependence of (a,c,e,g) resistivity and (b,d,f,h) magnetization for $\gcio $. From left to right, the Cd concentration is $x$=0.07, 0.08, 0.09, and 0.12. The vertical broken line indicates the magnetic transition temperature.
}
\end{center}
\end{figure}

\subsection{Magnetotransport properties for hole-doped $R$=Gd compounds}
Figure 9 shows the magnetic field dependence of resistivity, Hall resistivity, and magnetization for various $\gcio $. For all compounds, the resistivity remarkably decreases with increasing field at 2 K. The reduction of resistivity becomes smaller as the temperature is elevated. Hall resistivity for $x$=0.07 shows non-monotonic behavior whereas the magnetization simply increases as discussed in the main text. Figure 10 shows the magnetic field dependence of experimentally-obtained Hall resistivity (circles) and the fitting curve (solid lines). The deviation of the data from the fitting curve is explicit at 2 K. This behavior fades out as temperature and $x$ increases. For $x$=0.12, which is paramagnetic metal down to 2 K, the Hall resistivity seems to be proportional to the magnetization in the whole temperature and magnetic field region, as shown in Figs. 9

\begin{figure}
\begin{center}
\includegraphics[width=3.5in,keepaspectratio=true]{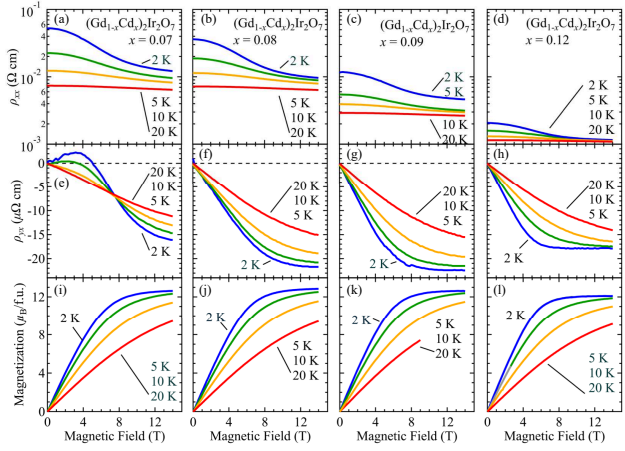}
\caption{(color online).
The magnetic field dependence of (a-d) Hall resistivity, (e-h) resistivity, (i-l) magnetization for $\gcio $. From left to right, the data are $x$=0.07, 0.08, 0.09, and 0.12.
}
\end{center}
\end{figure}

\begin{figure}
\begin{center}
\includegraphics[width=3.4in,keepaspectratio=true]{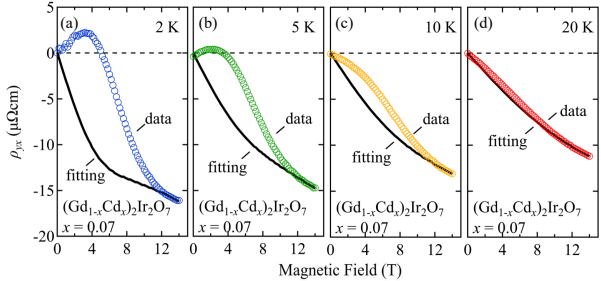}
\caption{(color online).
Magnetic field dependence of Hall resistivity for $\gcio $ ($x$=0.07) at several temperatures.
Thick black lines are fitting curves (see text).
}
\end{center}
\end{figure}


\begin{thebibliography}{100}

\bibitem{2018RMPArmitage} Armitage, N. P. and Mele, E. J. and Vishwanath, Ashvin, Rev. Mod. Phys. {\bf 90}, 015001 (2018).
\bibitem{2019NatureTokura} Y. Tokura, K. Yasuda, and A. Tsukazaki, Nature Review Physics, {\bf 1}, 126 (2019).
\bibitem{2011PRBWan} X. Wan, A. M. Turner, A. Vishwanath, S. Y. Savrasov, Phys. Rev. B {\bf 83}, 205101 (2011).
\bibitem{2011PRBZyuzin} Z. Zyuzin and A. A. Burkov, Phys. Rev. B {\bf 86}, 115133 (2012).
\bibitem{2010RMPNagaosa} N. Nagaosa, J. Sinova, S. Onoda, A. H. MacDonald, and N. P. Ong, Rev. Mod. Phys. {\bf 82}, 1539 (2010).
\bibitem{2015NatureNakatsuji} Satoru Nakatsuji, Naoki Kiyohara, and Tomoya Higo, Nature {\bf 527}, 212 (2015).
\bibitem{2016NPhysSuzuki} Suzuki, T., Chisnell, R., Devarakonda, A. Liu, Y. -T., Feng, W., Xiao, D., Lynn, J. W. and Checkelsky, J. G., Nature Phys. {\bf 12}, 1119 (2016).
\bibitem{2018NPhysLiu} Enke Liu {\etal}, Nature Phys. {\bf 14}, 1125 (2018).
\bibitem{2018NCommGhimire} Ghimire, Nirmal J., Botana, A. S., Jiang, J. S., Zhang, Junjie, Chen, Y. -S., and Mitchell, J. F., Nature Commun. {\bf 9}, 3280 (2018).
\bibitem{1998RMPImada} M. Imada, A. Fujimori, and Y. Tokura, Rev. Mod. Phys. {\bf 70}, 1039 (1998).
\bibitem{2010RMPGardner} J. S. Gardner, M. J. P. Gingras, and J. E. Greedan, Rev. Mod. Phys. {\bf 82}, 53 (2010).
\bibitem{2015NCommKondo} T. Kondo {\etal}, Nat. Commun. {\bf 6}, 10042 (2015).
\bibitem{2016PRBCano} J. Cano, B. Bradlyn, Z. Wang, M. Hirschberger, N. P. Ong, and B. A. Bernevig, Phys. Rev. B {\bf 95}, 161306 (2017).
\bibitem{2013PRLMoon} E.-G. Moon, C. Xu, Y.-B. Kim, and L. Balents, Phys. Rev. Lett. {\bf 111}, 206401 (2013). 
\bibitem{2012PRBKrempa} W. Witczak-Krempa, and Y.B. Kim, Phys. Rev. B {\bf 85}, 045124 (2012).
\bibitem{2010NatureMachida} Y. Machida, S. Nakatsuji, S. Onoda, T. Tayama, and T. Sakakibara, nature {\bf 463}, 2120 (2010). 
\bibitem{2018NCommUeda} K. Ueda, R. Kaneko, H. Ishizuka, J. Fujioka, N. Nagaosa, and Y. Tokura, Nat. Commun. {\bf 9}, 3032 (2018).
\bibitem{2015PRLUeda} K. Ueda, J. Fujioka, B.-J. Yang, J. Shiogai, A. Tsukazaki, S. Nakamura, S. Awaji, N. Nagaosa, and Y. Tokura, Phys. Rev. Lett. {\bf 115}, 056402 (2015).
\bibitem{2016NPhysTian} Z. Tian {\it et al.}, Nat. Phys. {\bf 12}, 134 (2016).
\bibitem{2017NCommUeda} K. Ueda, T. Oh, B.-J. Yang, R. Kaneko, J. Fujioka, N. Nagaosa, and Y. Tokura, Nat. Commun. {\bf 8}, 15515 (2017).
\bibitem{2014PRBUeda} K. Ueda, J. Fujioka, Y. Takahashi, T. Suzuki, S. Ishiwata, Y. Taguchi, M. Kawasaki, and Y. Tokura, Phys. Rev. B {\bf 89}, 075127 (2014).
\bibitem{2015ScienceMa} Eric Yue Ma, Yong-Tao Cui, Kentaro Ueda, Shujie Tang, Kai Chen, Nobumichi Tamura, Phillip M. Wu, J. Fujioka, Y. Tokura, and Zhi-Xun Shen, Science {\bf 350}, 538 (2015).
\bibitem{2016PRBUeda} K. Ueda, J. Fujioka, and Y. Tokura, Phys. Rev. B {\bf 93}, 245120 (2016).
\bibitem{2002JPSJFukazawa} H. Fukazawa and Y. Maeno, J. Phys. Soc. Jpn. {\bf 71}, 2578 (2002).
\bibitem{2019PRBKaneko} R. Kaneko, M.-T. Huebsch, S. Sakai, R. Arita, H. Shinaoka, K. Ueda, Y. Tokura, and J. Fujioka, Phys. Rev. B {\bf 99}, 161104 (2019).
\bibitem{2014PRBZhu} W. K. Zhu, M. Wang, B. Seradjeh, F. Yang, and S. X. Zhang, Phys. Rev. B {\bf 90}, 054419 (2014).
\bibitem{2019PRBPorter} Z. Porter, E. Zoghlin, S. Britner, S. Husremovic, J. P. C. Ruff, Y. Choi, D. Haskel, G. Laurita, and S. D. Wilson, Phys. Rev. B {\bf 100}, 054409 (2019).
\bibitem{2012PRLUeda} K. Ueda, J. Fujioka, Y. Takahashi, T. Suzuki, S. Ishiwata, Y. Taguchi, and Y. Tokura Phys. Rev. Lett. {\bf 109}, 136402 (2012).
\bibitem{2012PRBTafti} F. F. Tafti, J. J. Ishikawa, A. McCollam, S. Nakatsuji, and S. R. Julian, Phys. Rev. B {\bf 85}, 205104 (2012).
\bibitem{2012PRLUdagawa} M. Udagawa, H. Ishizuka, and Y. Motome, Phys. Rev. Lett {\bf 108}, 066406 (2012).
\bibitem{2016PRLWang} Z. Wang, K. Barros, G.-W. Chern, D. L. Maslov, and C. D. Batista, Phys. Rev. Lett {\bf 117}, 206601 (2016).
\bibitem{2003ScienceFang} Fang, Zhong and Nagaosa, Naoto and Takahashi, Kei S. and Asamitsu, Atsushi and Mathieu, Roland and Ogasawara, Takeshi and Yamada, Hiroyuki and Kawasaki, Masashi and Tokura, Yoshinori and Terakura, Kiyoyuki, Science {\bf 302}, 92 (2003).
\bibitem{2007PRLMiyasato} T. Miyasato, N. Abe, T. Fujii, A. Asamitsu, S. Onoda, Y. Onose, N. Nagaosa, and Y. Tokura, Phys. Rev. Lett. {\bf 99}, 086602 (2007).
\bibitem{2007PRLIguchi} S. Iguchi, N. Hanasaki, and Y. Tokura, Phys. Rev. Lett. {\bf 99}, 077202 (2007).
\bibitem{2018PRBOh} T. Oh, H. Ishizuka, and B.-J. Yang, Phys. Rev. B {\bf 98}, 144409 (2018).
\bibitem{2013PRBGuo} H. Guo, K. Matsuhira, I. Kawasaki, M. Wakeshima, Y. Hinatsu, I. Watanabe, and Z.-A. Xu, Phys. Rev. B {\bf 88}, 060411(R) (2013).
\bibitem{2008PRBOnoda} S. Onoda, N. Sugimoto,and N. Nagaosa, Phys. Rev. B {\bf 77}, 165103 (2008).
\bibitem{2020SciAdvKim} W. J. Kim, T. Oh, J. Song, E. K. Ko, Y. Li, J. Mun, B. Kim, J. Son, Z. Yang, Y. Kohama, M. Kim, B.-J. Yang, and T. W. Noh, Sci. Adv. {\bf 6}, 1539 (2020).
\bibitem{2017PRBGoswami} Pallab Goswami, Bitan Roy, and Sankar Das Sarma, Phys. Rev. B {\bf 95}, 085120 (2017).
\bibitem{2020arXivLadovrechis} Konstantinos Ladovrechis, Tobias Meng, and Bitan Roy, arXiv: 2012.10442 (2020).




\end{thebibliography}
\end{document}